\documentclass{mem}
\usepackage{natbib}\usepackage{txfonts}\usepackage{balance}
\usepackage{graphicx}
\usepackage[a4paper,breaklinks,dvipdfm]{hyperref}
\idline{75}{282}
\begin{document}
\def\teff{$T\rm_{eff }$}
\def\kms{$\mathrm {km s}^{-1}$}

\title{
Globular clusters: a chemical roadmap between anomalies and homogeneity}

   \subtitle{}

\author{
A. \,Mucciarelli\inst{1} 
   }

  \offprints{A. Mucciarelli}

\institute{
Dipartimento di Fisica \& Astronomia --
Universit\'a degli Studi di Bologna, Viale Berti Pichat 6/2,
I-40127 Bologna, Italy
\email{alessio.mucciarelli2@unibo.it}
}

\authorrunning{Mucciarelli}

\titlerunning{Globular clusters}

\abstract{

For several decades, globular clusters have been considered the best example of simple 
stellar populations, hosting coeval and chemical homogeneous stars. 
The last decade of spectroscopic and photometric studies has revealed a more complex 
view of their chemical composition, with a high level of homogeneity in their iron content 
but star-to-star variations in some light elements. This contribution summarizes  
the main evidence about the chemical anomalies in the stellar 
content of the globular clusters, discussing also some peculiar objects with 
intrinsic dispersions in their iron content.

}
\maketitle{}

\section{Introduction}

Globular clusters (GCs) are usually considered as the best example available 
in the nature of simple stellar populations (SSPs), aggregates of stars 
with the same age and initial chemical composition 
\citep[see e.g. the seminal paper by][]{rb86}. 
GCs as SSPs are valuable tools to approach 
several aspects of the astrophysics: they allow to check the predictions 
of the stellar evolution theory, to study the chemical enrichment history of 
the parent galaxy, 
and to investigate the properties of unresolved stellar populations (because 
for GCs in the Milky Way and in nearby galaxies we have simultaneously 
the resolved and integrated information). 

A large number of photometric and spectroscopic evidence collected in the last decades 
has conclusively established that the stars in a GC do not share the 
same initial chemical composition. 
In fact: 
(i)~some elements have been observed to vary in all the GCs, like 
C, N, O, Na and Al;
(ii)~some elements have been observed to vary only in some GCs, like 
He, Li, Mg, Si, K;
(iii)~finally, there is a bunch of {\sl strange beasts} characterized by 
a more or less pronounced Fe star-to-star variations.

As a general {\sl golden rule} we can define as genuine GCs all the massive, stellar 
systems that are homogeneous in their Fe (and Fe-peak) content. 
Generally, all the stars in a GC have the same iron content. Iron is 
produced by both Type II and Type Ia Supernovae (SN). Thus, 
stellar systems with an intrinsic Fe dispersion in their stellar content 
did retain the SN ejecta in their gravitational well, while systems like 
the genuine GCs did not retain these ejecta.

\citet{willman} performed a comparison between the intrinsic 
Fe spread (calculated by taking into account the uncertainties in each
individual star) of GCs and other stellar systems, like dwarf spheroidal
galaxies and ultra-faint dwarfs, finding that the GCs are characterized by 
intrinsic Fe spreads smaller than 0.05 dex, while dwarf systems have 
Fe spread larger than 0.2 dex. Thus, the level of homogeneity in the Fe content 
represents the main chemical fingerprint to distinguish GCs from other (more
complex) stellar systems.
Obviously, this definition does not allow to distinguish between GCs and 
less massive stellar systems without Fe spread, like the open clusters.
In this context, it is important to recall that all the GCs properly observed so far with 
large samples of high-resolution spectra exhibit also star-to-star variations 
among the light elements (C, N, O, Na, Mg and Al), known since forty years ago.
Even if the precise mechanism able to produce this kind of chemical variations 
are still debated and under scrutiny (see Section~\ref{lig}), these 
chemical anomalies are observed only in GCs (and not in open clusters or in field stars), 
and they are considered as typical feature of the GCs.

\section{Strange beasts}
\label{sb}

Up to now, only 5 (out of $\sim$150) Galactic GCs exhibit
intrinsic Fe spreads, namely Omega Centauri, Terzan 5, M54, M22 and NGC1851. 
Throughout this contribution, I refer only to the GCs for which chemical abundances 
from high-resolution spectroscopy and large sample of stars are available.
Other 2 GCs have been proposed to have a Fe spread, namely NGC~5824 and NGC~3201. 
The analysis of NGC~5824 presented by \citet{saviane12} 
is based on the Ca II triplet (as a proxy of the metallicity) but chemical analysis 
based on the direct measurement of Fe lines are still lacking. 
Recently, \citet{simmerer13} found an appreciable spread among the stars of NGC~3201, 
by using high-resolution spectra of 24 giants. However, other analysis always based on high-resolution 
spectra \citep{carretta09, munoz13} seem to contradict this finding, without 
clear hints of intrinsic Fe spreads. 
The normalized metallicity distributions of the 5 {\sl beasts} are shown in Fig.~1 
and briefly explained as follows:

(1)~Omega Centauri --- 
The most famous case of GC-like system with a intrinsic Fe variation is Omega Centauri, 
whose metallicity distribution has been investigated by several authors 
\citep[see e.g.][]{freeman,pancino02,johnson10,marino_omega}. 
Its metallicity distribution is very large (covering up to 1.5 dex) and 
multi-modal, with at least 5 peaks. However, the most popular scenario is that Omega Centauri 
is not a genuine GC but it is the remnant core of a tidally disrupted dwarf galaxy.

(2)~Terzan 5 --- 
Another case of mult-modal and broad Fe distribution is Terzan~5, 
a GC located in the inner bulge and known to harbor two red clumps in its 
color-magnitude diagram \citep{f09}. These two red clumps are associated 
to two different Fe abundances at about [Fe/H]=--0.30 and [Fe/H]=+0.30 dex \citep{f09}. 
The same Fe difference has been observed among the giant 
stars \citep{origlia11}, with the detection of an additional, third component 
at [Fe/H]$\sim$--0.8 dex \citep{origlia13}. 
The metallicity distribution of Terzan 5 turns out to be very 
large (about 1.5 dex) with at least 3 distinct peaks 
(see also the contribution by D. Massari to this conference).  
The striking chemical similarity between Terzan 5 and the bulge 
seems to suggest that Terzan 5 could be the 
remnant of one of the pristine fragment that contributed to form the Galactic bulge 
\citep{f09}.

(3)~M54 ---
A different case is provided by M54, a massive GC immersed in the nucleus of 
the Sagittarius remnant. \citet{carretta_m54} analyzed with FLAMES@VLT 
76 stars of M54, finding a broad ($\sim$0.7 dex) but 
uni-modal distribution.

(4)~M22 --- 
This cluster has been suspected to harbor an intrinsic Fe 
spread since thirty years ago \citep{pilachowski} even if other analysis 
have ruled out such a variation \citep{cohen,gratton82}. 
Recently, \citet{marino09} and \citet{marino11} performed the analysis of
high-resolution spectra for a total of 35 stars of M22, finding a broad but unimodal
metallicity distribution (ranging from [Fe/H]$\sim$--2 dex to $\sim$--1.6 dex). 
Also, M22 is known to harbor a split at the level of the sub-giant branch, with 
two distinct branches associated to different abundances of s-process elements \citep{marino12}.

(5)~NGC1851 --- 
Based on the analysis of 124 stars, \citet{carretta_1851} claim a small intrinsic 
Fe dispersion for this cluster.
When the metallicity distribution is roughly divided in two groups, the 
{\sl metal-poor} stars turn out to be more centrally concentrated with respect to 
the {\sl metal-rich} ones, suggesting a kinematical difference between the two groups 
of stars \citep[in fact,][proposed a merging as possible origin of NGC~1851]{carretta_1851}.
Note that other works did not found hints of intrinsic inhomogeneity 
\citep{yong08,vill10} even if based on smaller samples. Also, the analysis by \citet{willman} 
suggests that the observed Fe spread is fully compatible with those in other GCs.
Like M22, also NGC~1851 exhibits a split in its sub-giant branch, likely 
explainable with a difference in the total C+N+O.
The C+N+O in NGC~1851 is still a controversial and open issue.
\citet{yong08} measured a spread in the C+N+O content of about 0.6 dex 
among 4 bright giants in NGC~1851. These results have not been confirmed 
by the analysis of 15 giants by
\citet{vill10}, finding no differences in their C+N+O.

\begin{figure*}[h!]
\resizebox{\hsize}{!}{\includegraphics[clip=true]{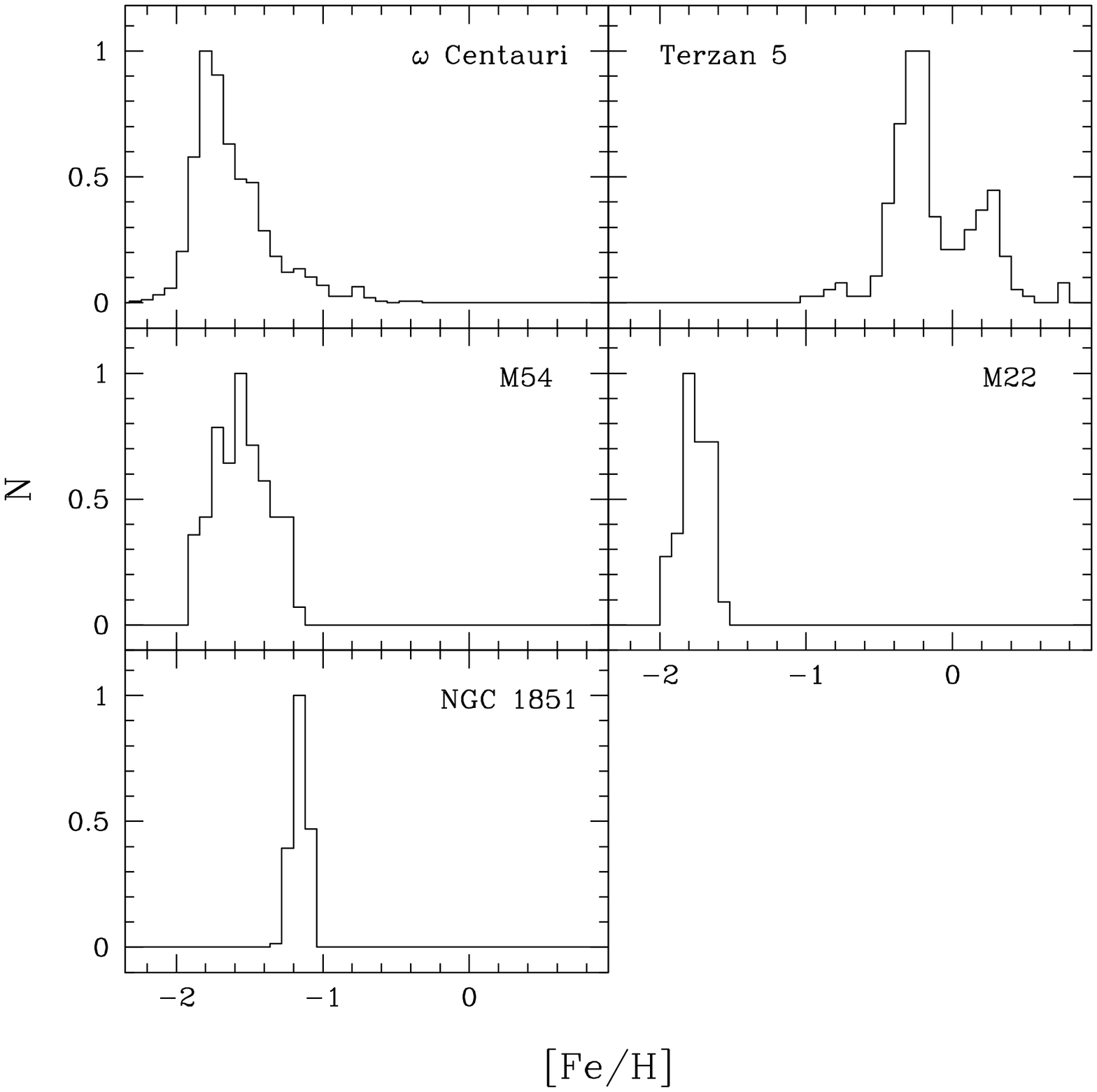}}
\caption{\footnotesize 
Normalized metallicity distribution for the 5 GC-like systems with an
intrinsic Fe spread: Omega Centauri \citep{johnson10}, 
Terzan 5 (Massari et al., in prep.), M54 \citep{carretta_m54}, M22 \citep{marino09,marino11} 
and NGC~1851 \citep{carretta_1851}.
}
\label{spread}
\end{figure*}

\section{Elements that vary}
\label{ele}

\subsection{C, N, O, Na, Mg, Al}
\label{lig}

The first evidence of inhomogeneity in the chemical content 
of GCs has been provided from the study of CN and CH features in the brightest giants 
by using low-resolution spectroscopy. \citet{osborn71} 
detected for the first time 2 CN-strong stars in M5 and M10, and following studies 
\citep{norris81,smith82,briley93,martell09,pancino10} have revealed that the GCs 
show a bimodality in their CN absorption and anti-correlation between CN and CH strengths, 
both among giant and dwarf stars. CN-strong/CH-weak stars, observed only in GCs, 
can be interpreted 
as having some amount of CNO processed material in their atmospheres.
The use of high-resolution spectroscopy for giant stars in GCs has allowed to link 
these CN/CH anomalies to other chemical anomalies. CN strength is correlated 
with Na and Al \citep{cottrell81} and anti-correlated with O \citep{sneden92}. 
Also, anti-correlations between O and Na exist, as discovered by the Lick-Texas group 
\citep[see Fig.~16 in][for a summary]{ivans01}. 

Currently, the Na-O anti-correlation is recognized as a typical feature of all the old and massive GCs, 
as widely demonstrated by the homogeneous survey of more than 1000 giants in 19 GCs 
performed by \citet{carretta09}. These anomalies have been observed also in old, massive 
extra-galactic GCs, like in Fornax \citep{letarte06} and in the Large Magellanic Cloud 
\citep{mucciarelli09}. 

A similar feature, often observed in some GCs, is the anti-correlation between Al and Mg.
%However, from a qualitative point of view, the definition of Mg-Al anti-correlation could be 
%incorrect. 
The label of "anti-correlation" could be improper, because 
most of the GCs show a large Al spread coupled with an unique value of Mg 
(in a given cluster), while the Mg-poor stars seem to be very rare. 
Up to now, GC stars with [Mg/Fe]$<$0 have been detected only in NGC~2808 \citep{carretta09}, 
in NGC~1786 \citep{mucciarelli09} and NGC~2419 \citep[][]{mucciarelli_2419}.

All the spectroscopic evidence collected so far 
(i.e. CN-CH anti-correlation, CN bimodality, Na-O anti-correlation, Mg-Al anti-correlation) 
are commonly interpreted as the signature of material processed through the high temperature 
extension of the proton-capture reactions (like NeNa and MgAl cycles).
Several theoretical models have been proposed in order to describe the formation and 
early evolution of GCs, for instance \citet{dercole08}, \citet{decressin10}, \citet{bekki11},
\citet{conroy11} and \citet{valcarce11}.
However, the basic idea behind these models is that the clusters formed with a chemical 
composition that well resembles that observed in the field stars. Then, they experience 
a period of star formation activity, with new stars born from a gas polluted by ejecta 
from the first generation stars that have processed material through the hot H-burning. 
Different kinds of polluter stars  
have been proposed, i.e. the asymptotic giant branch (AGB) stars and the 
fast-rotating massive stars. In both cases, a short ($<$100 Myr) 
phase of star-formation should occur in the early stage 
of the life of GCs.

Generally speaking, we used to separate the stars in a GC in two classes, a 
{\sl first generation} including stars with chemical abundances similar to those 
observed in field stars of similar metallicity (thus, Na-poor/O-rich/CN-weak stars), 
and a {\sl second generation} including Na-rich/O-poor/CN-strong stars. 
Obviously, this is a brutal but efficient classification. Considering our current knowledge and 
understanding of the GCs, we cannot exclude that the star-formation in a GC occurs in a continuos way 
and not with discrete bursts (as the classification in first and second generations seems to suggest).
This scheme is contradicted by the Na-O anticorrelation, 
that appears to be continuous \citep[with the only exception of M4, see][]{marino_m4}, 
while other findings (like the CN-bimodability and the red giant branch splitting observed 
in the color-magnitude diagrams including the U-band filter) seem to support this 
scenario.

Note that none of the current models are able to fully 
reproduce the observed chemical patterns, requiring some fine tuning in most of the 
main parameters.
Currently, a number of issues remain open and unsolved, like the nature of the polluters, 
the need of dilution of polluting gas with unprocessed material or the relative fraction 
between first and second generation stars.

\subsection{Helium}

The CN-strong and Na-rich/O-poor stars are expected to be also enriched in helium, 
whatever the polluter stars are, being helium the main product of the H-burning. 
The occurrence of (mild or strong) enhancement in Y (up to Y$\sim$0.4) has been 
proposed as the cause of  the  main sequence splitting 
and the peculiar horizontal branch morphology in the cases of Omega Centauri \citep{bedin_omega}, 
NGC~2808 \citep{piotto07,dalessandro}, NGC~2419 \citep{dicri} and NGC~6397 \citep{milone}.

The direct measurement of the He abundance in GC stars 
is a quite hard task, first of all because of the small number of available 
He diagnostics. The chromospheric line at 10820 $\AA$ can be 
observed in giant stars but it needs very high signal-to-noise and resolution 
spectra, and it is also heavily sensitive to the adopted modeling 
of the stellar chromosphere. This line has been used to infer the He 
abundance in giant stars of Omega Centauri \citep{dupree11,dupree13}
and NGC~2808 \citep{pasquini11}. The latter provided a differential 
analysis between two giant stars of NGC~2808 with different 
Na abundances. Their analysis points out a difference in Y between the two 
stars of at least 0.17, with the Na-rich stars being more He enriched than the 
Na-poor ones. 

Only a few photospheric lines can be detected among the horizontal branch stars, 
with effective temperatures between $\sim$9000 K and $\sim$12000 K \citep[the latter 
corresponding to the {\sl Grundahl Jump,}][]{grund}. 
\citet{villanova09} performed a chemical analysis in 4 horizontal branch stars 
in NGC~6752, finding an average value $<$Y$>$=~0.25, with values of [O/Fe] and 
[Na/Fe] compatible with those observed in the so-called first cluster generation. 
Also, the analysis of 6 HB stars in M4 by \citet{villanova12_m4} provides a higher 
He abundance, with an average value $<$Y$>$=~0.29, coupled with high values of [Na/Fe] 
and low values of [O/Fe]. 
Recently, \citet{marino_2808} measured the He abundance in 17 horizontal branch stars of NGC~2808, 
finding an average value of $<$Y$>$=~0.34.
These first findings strongly support the scenario where second generation stars are also 
enriched in He.

\subsection{Lithium}
Lithium remains a key-element in the study of the formation and evolution of GCs
but also an unresolved riddle. It is destroyed at temperatures of about 
2.5$\cdot10^6$ K, thus it cannot survive at the typical temperatures of the hot H-burning, 
larger than $\sim$10$^7$ K. A reasonable expectation is
that the second generation stars should be depleted in Li. Also, some kind of correlations/anti-correlations 
between Li and the elements involved in the chemical anomalies are expected, in particular 
Li-Na anticorrelations and Li-O correlations. Thus, the Li abundance 
could represent a formidable {\sl smoking gun} to disentangle the nature of the polluter stars 
able to create the second generation stars. 

The observational evidence collected so far provides us a scenario 
not easy to explain, giving us more questions and doubts than answers. 
Some GCs show clear or weak hints of Li spread and anti/correlations 
with Na or O, like 47 Tucanae \citep{bonifacio07_li}, 
NGC~6752 \citep{shen10_li}, NGC~6397 \citep{lind09_li} and M4 
\citep{monaco12_li}. On the other hand, other works seem to rule out 
Li spreads, like in M4 \citep{dorazi10_lia,dorazi10_lib,mucciarelli11_li}. 
%Note that in the case of M4, the results available in the literature are 
%not totally incompatible each other: \citet{dorazi10_lia} and \citet{mucciarelli11_li} 
%found no correlation between Li and O both among dwarf and giant stars, while 
%\citet{monaco12_li} detected a shallow slope between Li and Na. 

However, the small variation of Li associated in some cases to large variations of Na and/or O 
remains an open issue, and  the current theoretical models are not able to 
reproduce this finding. The models where the chemical anomalies are driven 
by AGB stars are able to explain a Li production through the Cameron-Fowler 
mechanism, but it needs a high degree of fine-tuning, in order to produce 
exactly the same amount of Li previously destroyed. The models based on 
fast-rotating massive stars do not include Li production mechanisms, 
invoking dilution processes to partially explain the very similar Li abundances 
between first and second generations.

%Thus, no firm conclusion can be drawn (at least today) from the Li abundances

\subsection{Potassium}
Potassium is a {\sl new entry} among the elements observed to vary only in some GCs.
In fact, the only cluster observed so far harboring a large dispersion of K 
is NGC~2419 \citep{mucciarelli_2419,cohen12}. This cluster shows two distinct groups 
of stars, the first characterized by {\sl normal} values of [Mg/Fe] and [K/Fe] 
(compatible with those observed in other GCs), 
the second with extreme values for both the abundance ratios, reaching 
[Mg/Fe]$\sim$--1.4 dex and [K/Fe]$\sim$+2.0 dex. 
These values are unusual for GC stars but these Mg-poor/K-rich stars represent 
$\sim$40\% of the studied samples.
Also, a Mg-K anti-correlation is clearly detected, suggesting 
that these anomalies can be explained within the self-enrichment scenario.
Interestingly enough, the fraction of Mg-poor stars well resembles that proposed 
by \citet{dicri} for the extreme He-rich population with Y=~0.4, according 
to the morphology of the horizontal branch of NGC~2419.

%A prompt theoretical framework has been proposed by 
\citet{ventura12} identified the Mg-poor, K-rich stars as the signature of an extreme nucleosynthesis 
driven by AGB and super-AGB stars, because K can be produced by proton capture on 
Argon nuclei during the normal nuclear reactions that create the observed chemical anomalies.

The first investigations about possible K spreads among the stars of other GCs 
have not shown chemical patterns similar to those observed in NGC~2419 
\citep{carretta_k}.

\section{Some thoughts about globular clusters}

The deep investigations about the chemical composition in the GCs performed 
in the last decade (and gathered by the massive use of high-resolution 
spectrographs) has for some aspects changed our view of the GCs, unveiling 
a quite complex formation process.

Only to provide a simple, mental scheme to put some order among this 
evidence, we can divide the GCs in three classes: 
(1)~the genuine GCs, characterized by homogeneity in their Fe content, 
that show a uni-modal and narrow metallicity distribution;
(2)~the GCs with large but uni-modal metallicity distribution, like M54, M22 
and NGC1851, even if an enlarge of their studied samples is mandatory in order 
to unveil possible secondary peaks;
(3)~a third group of systems, including only Omega Centauri and Terzan~5 with 
broad and multi-modal metallicity distribution, that cannot be considered 
genuine GCs, but stellar systems undergoing complex chemical enrichment
histories.

Can we continue to use genuine GCs as simple stellar populations, 
in spite of their chemical anomalies and the proposed self-enrichment 
scenario? A reasonable answer is yes (but with some cautions).
In fact:
(1)~They are homogeneous in Fe and almost any other elements, 
with the exceptions of C, N, O, Na, Mg and Al. Thus we can continue to 
use them as tracers of the chemical composition of the host galaxy, 
by focusing our attention on the elements (the majority) that do not 
show intrinsic star-to-star variations.
(2)~They are still single-age stellar systems. The star formation in the GCs 
occurs within short timescales ($<$ 100 Myr). This timescale 
is much smaller than the GC age and the age differences between 
GC sub-populations with different chemical compositions (in terms of light 
elements) cannot be appreciated at the level of the turnoff. 
(3)~Their integrated colors can be still used to calibrate and study 
unresolved stellar populations, like the GCs outside the Local Group, because 
the effect of the chemical anomalies on the integrated colors is negligible 
\citep[with the only caution to exclude colors including U filters, heavily affected 
by C and N variations, see][]{sbordone}.

Are the GCs strictly speaking simple stellar populations?
The answer is no, but they have never been considered so. %On the other hand, 
Are the GCs the simplest simple stellar populations 
available in the Universe? The answer is clearly yes, whatever complex their 
formation scenario is.

\begin{acknowledgements}
This research is part of the project COSMIC-LAB 
(web site: www.cosmic-lab.eu) funded by the European Research 
Council (under contract ERC-2010-AdG-267675).
\end{acknowledgements}

\bibliographystyle{aa}

\begin{thebibliography}{}

\bibitem[Bedin et al.(2004)]{bedin_omega}
Bedin, R. L., Piotto, G., Anderson, J., Cassisi, S., King, I. R., 
Momany, Y., \& Carraro, G., 2004, ApJ, 605L, 125

\bibitem[Bekki et al.(2011)]{bekki11}
Bekki, K., 2011, MNRAS, 412, 2241

\bibitem[Briley et al.(1993)]{briley93}
Briley, M. M.., Smith, G. H., Hesser, J. E., \& Bell, R. A., 1993, AJ, 106, 142

\bibitem[Bonifacio et al.(2007)]{bonifacio07_li}
Bonifacio, P., Pasquini, L., Molaro, P., Carretta, E., Francois, P., 
Gratton, R. G., James, G., Sbordone, L., Spite, F., \& Zoccali, M., 
2007, A\&A, 470 153

\bibitem[Carretta et al.(2009)]{carretta09}
Carretta, E. et al., 2009, A\&A, 505, 117

\bibitem[Carretta et al.(2010a)]{carretta_m54}
Carretta, E., et al., 2010, A\&A, 520, 95

\bibitem[Carretta et al.(2010b)]{carretta_1851}
Carretta, E., et al., 2010, ApJ, 722L, 1

\bibitem[Carretta et al.(2013)]{carretta_k}
Carretta, E., Gratton, R. G., Bragaglia, A., D'Orazi, V., 
Lucatello, S., Sollima, A., \& Sneden, C., 2013, ApJ, 769, 40

\bibitem[Cohen(1981)]{cohen}
Cohen, J. G., 1981, ApJ, 247, 869

\bibitem[Cohen \& Kirby(2012)]{cohen12}
Cohen, J. G., \& Kirby, E. N., 2012, ApJ, 760, 86

\bibitem[Conroy \& Spergel(2011)]{conroy11}
Conroy, C., \& Spergel, D. N., 2011, ApJ, 726, 36

\bibitem[Cottrell \& Da Costa(1981)]{cottrell81}
Cottrell, P. L., \& Da Costa, G. S., 1981, ApJ, 245L, 79

\bibitem[Dalessandro et al.(2011)]{dalessandro}
Dalessandro, E., Salaris, M., Ferraro, F. R., Cassisi, S., Lanzoni, B., 
Rood, R. T., Fusi Pecci, F., \& Sabbi, E., 2011, MNRAS, 410, 694

\bibitem[Decressin et al.(2010)]{decressin10}
Decressin, T., Baumgardt, H., Charbonnel, C., \& Kroupa, P., 2010, 
A\&A, 516, 73

\bibitem[D'Ercole et al.(2008)]{dercole08}
D'Ercole, A., Vesperini, E., D'Antona, F., McMillan, S. L. W., \& Recchi, S., 
2008, MNRAS, 391, 825

\bibitem[di Criscienzo et al.(2011)]{dicri}
di Criscienzo, M., D'Antona, F., Milone, A. P., Ventura, P., Caloi, V., 
Carini, R., D'Ercole, A., Vesperini, E. \& Piotto, G., 2011, MNRAS, 414, 3381

\bibitem[D'Orazi \& Marino(2010)]{dorazi10_lia}
D'Orazi, V., \& Marino, A. F., 2010, ApJ, 716L, 166

\bibitem[D'Orazi et al.(2010)]{dorazi10_lib}
D'Orazi, V., Lucatello, S., Gratton, R. G., Bragaglia, A., \& 
Carretta, E., 2010, ApJ, 713L, 1

\bibitem[Dupree et al.(2011)]{dupree11}
Dupree, A. K. , Strader, J., \& Smith, G. H., 2011, ApJ, 728, 155

\bibitem[Dupree \& Avrett(2013)]{dupree13}
Dupree, A. K., \& Avrett, E. H., 2013, ApJ, 773L, 28

\bibitem[Ferraro et al.(2009)]{f09}
Ferraro, F. R., Dalessandro, E., Mucciarelli, A., Beccari, G., 
Rich, R. M., Origlia, L., Lanzoni, B., Rood, R. T., Valenti, E., 
Bellazzini, M., Ransom, S. M., \& Cocozza, G., 2009, Nature, 462, 483

\bibitem[Freeman \& Rodgers(1975)]{freeman}
Freeman, K. C., \& Rodgers, A. W., 1975, ApJ, 201L, 71

\bibitem[Gratton(1982)]{gratton82}
Gratton, R. G., 1982, A\&A, 115, 171

\bibitem[Grundahl et al.(1999)]{grund}
Grundahl, F., Catelan, M., Landsman, W. B., Stetson, P. B., \& Andersen, M. I., 
1999, ApJ, 524, 242


\bibitem[Johnson \& Pilachowski(2010)]{johnson10}
Johnson, C. I., \& Pilachowski, C. A., 2010, ApJ, 722, 1373


\bibitem[Ivans et al.(2001)]{ivans01}
Ivans. I. I., Kraft, R. P., Sneden, C., Smith, G. H., Rich, R. M., \& Shetrone, M., 2001, 
AJ, 122, 1438

\bibitem[Letarte et al.(2006)]{letarte06}
Letarte, B., Hill, V., Jablonka, P., Tolstoy, E., Francois, P., \& Meylan, G., 
2006, A\&A, 453, 547

\bibitem[Lind et al.(2009)]{lind09_li}
Lind, K. Primas, F., Charbonnel, C., Grundahl, F., \& Asplund, M., 
2009, A\&A, 503, 545

\bibitem[Marino et al.(2011)]{marino_omega}
Marino, A. F., et al., 2011, ApJ, 731, 64

\bibitem[Martell \& Smith(2009)]{martell09}
Martell, S. L., \& Smith, G. H., 2009, PASP, 121, 577

\bibitem[Milone et al.(2012)]{milone}
Milone, A. P., Marino, A. F., Piotto, G., Bedin, L. R., Anderson, J., Aparicio, A., 
Cassisi, S., \& Rich, R. M., 2012, ApJ, 745, 27

\bibitem[Marino et al.(2008)]{marino_m4}
Marino, A. F., Villanova, S., Piotto, G., Milone, A. P., Momany, Y., Bedin, L. R., 
\& Medling, A. M., 2008, A\&A, 490, 625

\bibitem[Marino et al.(2009)]{marino09}
Marino, A. F., Milone, A. P., Piotto, G., Villanova, S., Bedin, L. R., 
Bellini, A., \& Renzini, A., 2009, A\&A, 505, 1099

\bibitem[Marino et al.(2011)]{marino11}
Marino, A. F., et al., 2011, A\&A, 532, 8

\bibitem[Marino et al.(2012)]{marino12}
Marino, A. F., et al., 2012, A\&A, 541, 15

\bibitem[Marino et al.(2013)]{marino_2808}
Marino, A. F., et al., 2013, MNRAS.tmp.2693

\bibitem[Monaco et al.(2012)]{monaco12_li}
Monaco, L., Villanova, S., Bonifacio, P., Caffau, E., Geisler, D., 
Marconi, G., Momany, Y., \& Ludwig, H.-G., 2012, A\&A, 539, 157


\bibitem[Mucciarelli et al.(2009)]{mucciarelli09}
Mucciarelli, A., Origlia, L., Ferraro, F. R., \& Pancino, E., 2009, ApJ, 695, 134


\bibitem[Mucciarelli et al.(2011)]{mucciarelli11_li}
Mucciarelli, A., Salaris, M., Lovisi, L., Ferraro, F. R., Lanzoni, B., 
Lucatello, S., \& Gratton, R. G., 2011, MNRAS, 412, 81


\bibitem[Mucciarelli et al.(2012)]{mucciarelli_2419}
Mucciarelli, A., Bellazzini, M., Ibata, R., Merle, T., Chapman, S. C., Dalessandro, E., 
\& Sollima, A., 2012, MNRAS, 426. 2889

\bibitem[Munoz et al.(2013)]{munoz13}
Munoz, C., Geisler, D., \& Villanova, S., 2013, MNRAS, 433, 2006


\bibitem[Norris et al.(1981)]{norris81}
Norris, J., Cottrell, P. L., Freeman, K. C., \& Da Costa, G. S., 1981, ApJ 244, 205

\bibitem[Origlia et al.(2011)]{origlia11}
Origlia, L., Rich, R. M., Ferraro, F. R., Lanzoni, B., Bellazzini, M., 
Dalessandro, E., Mucciarelli, A., Valenti, E., \& Beccari, G., 2011, 
ApJ, 726L, 20

\bibitem[Origlia et al.(2013)]{origlia13}
Origlia, L., Massari, D., Rich, R. M., Mucciarelli, A., Ferraro, F. R., Dalessandro, E., 
\& Lanzoni, B., 2013, ApJ, 779L, 5 


\bibitem[Osborn(1971)]{osborn71}
Osborn, W., 1971, Obs, 91, 223

\bibitem[Pancino et a.(2002)]{pancino02}
Pancino, E., Pasquini, L., Hill, V., Ferraro, F. R., \& Bellazzini, M., 2002, ApJ, 568L, 101

\bibitem[Pancino et al.(2010)]{pancino10}
Pancino, E., Rejkuba, M., Zoccali, M., \& Carrera, R., 2010, A\&A, 524, 44

\bibitem[Pasquini et al.(2011)]{pasquini11}
Pasquini, L., Mauas, P., Kaufl, H. U., \& Cacciari, C., 2011, A\&A, 
531, 35

\bibitem[Pilachowski et al.(1982)]{pilachowski}
Pilachowski, C., Leep, E. M, Wallerstein, G., \& Peterson, R. C., 1982, ApJ, 263, 187

\bibitem[Piotto et al.(2007)]{piotto07}
Piotto, G., Bedin, L. R., Anderson, J., King, I. R., Cassisi, S., Milone, A. P., 
Villanova, S., Pietrinferni, A., \& Renzini, A., 2007, ApJ, 661L, 53

\bibitem[Renzini \& Buzzoni(1986)]{rb86}
Renzini, A. \& Buzzoni, A., 1986, Spectral Evolution of Galaxies, 122, 195

\bibitem[Saviane et al.(2012)]{saviane12}
Saviane, I., da Costa, G. S., Held, E. V., Sommariva, V., Gullieuszik, M., 
Barbuy, B., \& Ortolani, S., 2012, A\&A, 540, 27

\bibitem[Sbordone et al.(2011)]{sbordone}
Sbordone, L., Salaris, M., Weiss, A., \& Cassisi, S., 2011, A\&A, 534, 9

\bibitem[Shen et al.(2010)]{shen10_li}
Shen, Z.-X., Bonifacio, P., Pasquini, L., \& Zaggia, S., 2010, A\&A, 524L, 2

\bibitem[Simmerer et al.(2013)]{simmerer13}
Simmerer, J., Ivans, I. I., Filler, D., Francois, P., Charbonnel, C., Monier, R., 
\& James, G., 2013, ApJ, 764, 7L


\bibitem[Smith \& Norris(1982)]{smith82}
Smith, G. H., \& Norris, J., 1982, ApJ, 254, 594

\bibitem[Sneden et al.(1992)]{sneden92}
Sneden, C., Kraft, R. P., Prosser, C. F., \& Langer, G. E., 1992, AJ, 104, 2121

\bibitem[Valcarce \& Catelan(2011)]{valcarce11}
Valcarce, A. A. R., \& Catelan, M., 2011, A\&A, 533, 120

\bibitem[Ventura et al.(2012)]{ventura12}
Ventura, P., D'Antona, F., Di Criscienzo, M., Carini, R., D'Ercole, A., 
\& Vesperini, E., 2012, ApJ, 761L, 30

\bibitem[Villanova et al.(2009)]{villanova09}
Villanova, S., Piotto, G., \& Gratton, R. G., 2009, A\&A, 499, 755

\bibitem[Villanova et al.(2010)]{vill10}
Villanova, S., Geisler, D., \& Piotto, G., 2010, ApJ, 722L, 18

\bibitem[Villanova et al.(2012)]{villanova12_m4}
Villanova, S., Geisler, D., Piotto, G., \& Gratton, R. G., 2012, 
ApJ, 748, 62 

\bibitem[Yong \& Grundahl(2008)]{yong08}
Yong, D. \& Grundahl, F., 2008, ApJ, 672L, 29

\bibitem[Willman \& Strader(2012)]{willman}
Willman, B. \& Strader, J., 2012, AJ, 144, 76


\end{thebibliography}

\end{document}